\documentstyle[amssymb,epsf,graphics]{EuroPhys}
\input EuroMacr
\begin{document}
%
%
%
\euro{40}{5}{485-490}{1997}
\Date{1997}
\shorttitle{T.B. LIVERPOOL {\it et al}: SCREENED POLYELECTROLYTES.}
%
%
%
\title{The scaling behaviour of screened polyelectrolytes}
\author{T.B. Liverpool\inst{1}\footnote{email address:liverpoo@mpip-mainz.mpg.de} \And M. Stapper\inst{1}\footnote{email address:stapper@mpip-mainz.mpg.de}}
\institute{
     \inst{1} Max-Planck-Institut f\"ur Polymerforschung  
              Postfach 3148, D-55021 Mainz, Germany.\\}
%
%
\rec{}{}
%
%
%
\pacs{
\Pacs{05}{20$-$y}{}
\Pacs{36}{20$-$r}{}
\Pacs{64}{60$-$i}{}
      }
\maketitle
%
%
%
\begin{abstract}
  We present a field-theoretic renormalization group (RG) analysis of a single
  flexible, screened polyelectrolyte chain (a Debye-H\"uckel chain) in a polar
  solvent. We point out that the Debye-H\"uckel chain may be mapped onto a
  {\it local} field theory which has the same fixed point as a generalised
  $n\rightarrow1$ Potts model. Systematic analysis of the field theory shows
  that the system is one with two interplaying length-scales {\it requiring}
  the calculation of scaling functions {\it as well} as exponents to fully
  describe its physical behaviour. To illustrate this, we solve the RG
  equation and explicitly calculate the exponents and the mean end-to-end
  length of the chain.
\end{abstract}
%
%
%
%
%
Polyelectrolytes are polymers with ionizable monomers which in polar solvents,
such as water, dissociate into charged polymers and small `counter-ions' of
opposite sign. They are of widespread importance with many applications in
physics, biology and chemistry~\cite{Joanny}. Their solublility in polar
solvents means that they also have many industrial applications. Typical
examples are DNA and sulphonated polystyrene.

The behaviour of polyelectrolytes is quite poorly understood, chiefly because
it has been difficult to deal with their long-range coulomb interactions
theoretically. In addition, counter-ions, complex-formation and salt and their
differing length-scales make comparison between theory and experiment
difficult because of questions of validity of the theories in different
regimes. Neutral polymers on the other hand, have been extremely succesfully
described by scaling ideas and renormalization group theories~\cite{frdec}.
The separation of the length-scales in their physics leads to {\em
  universality}. The main properties of the system are independent of
microscopic details which only change prefactors in the physical quantities
and make analysis of experimental results, at least in their limiting
behaviour, reasonably straightforward. It would be particularly useful if the
behaviour of polyelectrolytes could also be put in a similar context. The
resulting theory would obviously be more complicated reflecting the increased
complexity of the system. Nonetheless a 'universal' theory of polylectrolytes
would be a real step forward in the understanding of polymer solutions. This
is what we attempt to do in this paper. In this spirit we consider the
simplest possible system, a single charged chain in solution.

In solution the counter-ions in the vicinity of the macro-ion(polymer)
`screen' or reduce the range of the coulombic interaction. The simplest model
of a screened polyelectrolyte which we use as the starting point for our
analysis is the Wiener chain with a Debye-H\"uckel (DH) or Yukawa potential
interaction between charged monomers. The chain is described by a vector ${\bf r}(s)$
parametrised by the arc-length $s$ and the partition function is obtained from
$Z=\int {\cal D}{\bf r}(s) \exp\left\{-\beta {\cal{H}}_{{DH}}[{\bf
    r}(s)]\right\}$ with the Hamiltonian of the system given in $d$ dimensions
by \begin{equation} \beta {\cal{H}}_{{DH}}= \frac{d}{2\ell} \int^{L}_{0} ds
  \left(\frac{\partial {\bf r}}{\partial s}\right)^{2}+
  b\int^{L}_{0}\int^{L}_{0} ds ds' V({\bf r}(s)-{\bf r}(s')),
  \label{dh}\end{equation} where $\ell$ is the `{\it Kuhn-segment}' length of
the chain, $L= \ell N$ the length of the chain, where $N$ is the number of
segments on the chain. The Kuhn-length here refers to the distance between
{\it charged} monomers (see inset Fig.1). $1/\beta=k_{B}T$ and $b \propto
\lambda_{B}$ where $\lambda_{B}=\frac{q^{2}}{4\pi\epsilon\epsilon_{0}k_{B}T}$
is the Bjerrum length which measures the strength of the bare coulomb
interaction. In water, $\lambda_{B}=7.14 \mbox{\AA}$. As usual
$\epsilon\epsilon_{0}$ is the dielectic constant of the solvent, $k_{B}$ is
the Boltzmann constant and $T$ the temperature. The potential between monomers,
$V({\bf r})$ is a solution of the linearised Poisson-Boltzmann equation
$(-\Delta + \kappa^{2})V({\bf r})=0$ which in three dimensions is given by the
usual $V({\bf r})={e^{-\kappa|{\bf r}|}}/{|{\bf r}|}$. The screening length or
range of the interaction is given by $1/\kappa$ and $\kappa \propto
\sqrt{\lambda_{B}}$ is a function of the density of screening ions and the
dielectric properties of the solution. The DH model~\cite{Vilgis} is thought
to be valid only when the polyelectrolyte is weakly charged (see inset Fig.1)
and in the presence of salt. There are no large fluctuations in the
counter-ion/salt density so for the Coulomb interaction the DH model may be
considered to be `{\it mean-field}'.

Recent extensive simulations of flexible DH chains~\cite{mickkrem} give
dramatically different results from existing theories~\cite{kappa2,kappa}. We
compare our results with these simulations because experiments are always
performed at non-zero concentrations where many-chain effects begin to play a
part and cannot be used to check the validity of a single chain
model~\cite{Joanny}.  The quantity that we want to calculate is the average
end-to-end length $\langle R^{2}\rangle \equiv \langle (r(L)-r(0))^{2}
\rangle$.

Returning to equation (\ref{dh}), we apply a generalisation of the well known
de Gennes trick to map the polyelectrolyte to an $n \rightarrow 0$ field
theory~\cite{deGennes72} with a non-local (long-ranged)
interaction~\cite{Pfeuty,Jug}. The propagator is defined in the usual way as
the Laplace transform (LT) of the 2-point function $G({\bf R})\equiv \int
[{\cal D}\phi]\phi({\bf R})\phi({\bf 0}) e^{-{ S(\phi)}} \nonumber
=\int_{0}^{\infty} d{\cal L} e^{-t{\cal L}}\langle\delta({\bf r}({\cal
  L})-{\bf r}(0)-{\bf R})\rangle $ where ${\cal L}=2 L \ell/d = 2 N\ell^{2}/d$
with the action given by $S(\phi)=\int_{\bf r}\left[ \frac{1}{2}(\nabla
  \phi)^{2}+ \frac{1}{2} t \phi^{2}\right]+ \frac{g}{2}\int_{\bf r}\int_{\bf
  r'}\phi^{2}({\bf r})V({\bf r}-{\bf r'})\phi^{2}({\bf r'})$ where
$2g=(2\ell/d)^{2} b$ and $\phi({\bf r})$ are $n \rightarrow 0$ component
fields with $O(n)$ symmetry.  We have used the notation $\int_{{\bf r}} \equiv
\int d^{d}r$. We perform a Hubbard-Stratonovich transformation~\cite{Pfeuty}
to remove the DH potential and generate a new scalar field $\psi({\bf r})$ so
that
\begin{eqnarray} &&G({\bf R})= \int 
  \frac{d^{d}q}{(2\pi)^{d}}e^{{i{\bf q}\cdot {\bf r}}}\tilde{G}(q)={\cal
    N}\int [{\cal D}\phi][{\cal D}\psi]\phi({\bf R})\phi({\bf
    0}) e^{-{S(\phi,\psi)}}  \\
  &&\mbox{with} \, \, \, S(\phi,\psi)= 
  \int_{\bf r}\left[ \frac{1}{2}(\nabla \phi)^{2}+\frac{1}{2} t_{2}^{(0)}
    \phi^{2}+\frac{1}{2}(\nabla \psi)^{2}+\frac{1}{2} t_{1}^{(0)} \psi^{2}+
    \frac{1}{3!}u_{0} \phi^{2}\psi\right] \label{bo}
\end{eqnarray} where $t_{2}^{(0)}$ is the polyelectrolyte fugacity, 
$t_{1}^{(0)}=\kappa^{2}$ and ${\cal N}$ is a normalisation factor.  The
coupling is given by $u_{0}/3!=i\sqrt{g}(\sqrt{g})$ if $\psi$ is
real(imaginary). $\psi({\bf r})$ is a fluctuating Coulomb field
which is coupled to the polymer density. It has fluctuations on a length-scale
given by $1/\kappa$. Since there are two independent fields, we have two sets
of exponents $\nu_{\psi},\nu_{\phi};\eta_{\psi},\eta_{\phi}$. To distinguish
between the two we use a superscript $\psi$ for the Coulomb terms and a
superscript $\phi$ for the polymer terms.

There is a rich literature on the use of the RG and field theory to calculate
the properties of neutral polymers. On the other hand, these techniques have
not been applied successfully to charged polymers, polyelectrolytes. This was
due to a misconception that they could not be usefully applied to such systems
in the physical dimension, $d=3$. In the literature on field theory and
renormalization group theory of polyelectrolytes~\cite{Pfeuty,Jug,Mezard},
work has always been done on unscreened chains. This has led to an unfortunate
obscuring of the physics which we believe we have finally clarified. The
exponents $\nu_{\phi}$ can be calculated to all orders in $\epsilon=6-d$ (the
upper critical dimension $d_{c}=6$ can be inferred from the Lagrangian
(\ref{bo})) and is given by $\nu_{\phi}=2/(d-2)$~\cite{Jug}. It has always
been assumed that the end-to-end length is given by
\begin{equation} \langle R^{2} \rangle \sim \ell^{2} N^{{2\nu_{\phi}}}
\label{stretch}\end{equation}
and since $\nu_{\phi}=1$ in $d=4$, and the chain could not be stretched out
more than a rod , $\langle R^{2} \rangle \ngtr \ell^{2} N^{2}$, it was
said that due to overstretching of the Gaussian chain the theory breaks down
for $d<4$. It was therefore thought that field theoretic methods could not be
used on charged chains in the physical dimension $d=3$. We will show that
this analysis is incomplete. 

We have a number of new results in our letter. Firstly, we point out a
connection between a Potts model used to study percolation and the sol-gel
transition and a single polyelectrolyte chain~\cite{Janssen}.  Secondly, we
include the two {\it important} length-scales which are necessary for the
analysis of the system. They are the screening-length $1/\kappa$ and the
gaussian chain size $S=\langle R_{0}^{2}\rangle^{1/2}= \ell N^{1/2}$. A wide
range of systems can be studied beginning where the screening is very high
(SAW), $\kappa^{2} \ell^{2} N >> 1$ (the polymer field $\phi$ is more critical
than coulomb field $\psi$) and ending where it is very low (unscreened,
rod-like), $\kappa^{2} \ell^{2} N << 1$ ($\psi$ is more critical than $\phi$).
This is therefore a problem of cross-over~\cite{AmitGold}. The previous work
has focussed, in our opinion mistakenly, on the unscreened case by setting
$\kappa=0$. The exponent $\nu_{\phi}$ is actually calculated for the
intermediate regime $\kappa^{2} \sim 1/(\ell^{2} N) \rightarrow 0$ and {\it
  does not} apply to the unscreened (rod) regime where $\kappa \rightarrow 0
\, \, \mbox{\bf and} \, \, \kappa^{2} \ell^{2} N \rightarrow 0$.  We also find
that quantities like the end-to-end distance depend on negative powers of
$\kappa$.  This implies that the operator $\int_{r}\kappa^{2}\psi^{2}$ in eqn.
(\ref{bo}) is dangerously irrelevant~\cite{ft}. By setting $\kappa=0$ at the
start of the calculation, one loses sight of such subtleties.

We calculate in agreement with earlier work~\cite{Pfeuty,Jug} the exponents
\begin{equation}
\eta_{\phi}=-\frac{\epsilon}{4} -\frac{9\epsilon^{2}}{32}+O(\epsilon^{3}),  \,\,\, \eta_{\psi}=0,  \,\,\, \nu_{\psi}= \frac{1}{2},  \,\,\, \nu_{\phi}=\frac{2}{d-2}+O(\epsilon^{3}); \,\,\,
\label{exponents}\end{equation} where $\epsilon=6-d$. The exponents can
also be obtained from the percolation theory 
but at a different fixed point from the usual sol-gel
transition~\cite{Janssen}.  This leads naturally
to our final conclusion: these exponents correspond to a {\bf bi-critical}
fixed point (both $\phi,\psi$ are critical) where the screening length and the
chain size are comparable and at this fixed point the end-to-end length we
calculate as
\begin{equation} \langle R^{2} \rangle \sim \ell^{2} N^{{2\nu_{\phi}}}f(\bar{\kappa}^{1/\nu_{\phi}}N)
\end{equation}
where $\bar{\kappa}/\ell= \kappa$ and $f(x)$ is a function {\it singular} as
$x \rightarrow 0$. We {\it explicitly} calculate $f(x)=x^{\left(- 2\nu_{\phi}+
    1\right)}g(x)$ where
$g(x)=(1-3\epsilon\log(x)/(4x))[1-\epsilon(3\gamma/(4x)-5/24)][1-\epsilon(3/(4x)-1/2+x/8)
\exp{(x)} E_{1}(x)] + 0(\epsilon^{2})$. The exponential integral $E_{1}(z)$ is
defined by $E_{1}(z)=\int_{1}^{{\infty}}dt \exp(-zt)/t$ and $\gamma=0.57721$
is Euler's constant. The end-to-end length is given by \begin{equation}
  \langle R^{2} \rangle \sim \ell^{2}
  \frac{N}{\bar{\kappa}^{2}}\bar{\kappa}^{1/(\nu_{\phi})}g(\bar{\kappa}^{1/\nu_{\phi}}N)
  \sim \ell^{2} N^{2-1/(2\nu_{\phi})}
  g(\bar{\kappa}^{1/\nu_{\phi}}N)\label{rcool}\end{equation} (Recall that at
this fixed point $\kappa^{2} \sim 1/(\ell^{2} N) \rightarrow 0$). Earlier
work~\cite{Pfeuty,Jug} had always {\it implicitly} assumed that $f(x)$ was a
well behaved function in the limit $x \rightarrow 0 \, \, (\kappa \rightarrow
0)$.  We find $2\nu'_{1}=2-1/(2\nu_{\phi})$ determines the chain size and {\it
  not} $2\nu_{\phi}$. Also, because of $g$ the exponent will vary slightly
with $\kappa$~\cite{mickkrem}. There is consequently no over-stretching (at
this fixed-point, $u*^{2}=-27\epsilon/4$) for $d<4$ and equation
(\ref{stretch}) should be replaced by equation (\ref{rcool}).  It is important
to note that $\nu'_{1}=7/8 =0.875$ (intermediate between the SAW, $\nu=0.588$
and rod, $\nu=1.0$) in three dimensions. It is also clear from our analysis
that the lower critical dimension is $d=2$ where the theory breaks down.

In order to compare with earlier approaches we may `fit' our results from
equation (\ref{rcool}) to a worm-like chain model~\cite{osf}. The
electrostatic persistence length, $\bar{\ell}_{e}$, may be estimated from
$\bar{\ell}_{e}(1-e^{-N\ell/\bar{\ell}_{e}}) \propto
\bar{\kappa}^{\nu'_{2}}{g}(\bar{\kappa}^2N^4)$ with $\nu'_{2} \simeq -3/2$ in
$d=3$ showing no unique power law behaviour. This is in agreement with the
simulations of Micka and Kremer~\cite{mickkrem}.  The rod-like behaviour in
the long-range, low salt regime $1/\kappa^{2}>>\ell^{2}N$ and the
self-avoiding walk behaviour for the short-range, high salt regime,
$\ell^{2}N>>1/\kappa^{2}$~\cite{solutions} can be obtained using different
methods~\cite{Joanny}.

Now we perform a RG analysis of our model with Lagrangian given by equation
(\ref{bo}). Our goal is to calculate
\begin{equation}
\langle R^{2}\rangle= -{\cal
  T}^{-1}_{\cal L}(\partial_{q^{2}}\tilde{G}(q)|_{q=0})/{\cal
  T}^{-1}_{\cal L}(\tilde{G}(q)|_{q=0})=D(u)F({\cal L},t_{1}) \label{Radius2}
\end{equation} where $D(u)$ is a non universal constant, ${\cal  T}^{-1}_{\cal
  L}$ is the inverse LT with respect to ${\cal L}$ and
$\tilde{G}(q)\equiv\tilde{G}^{(2,0)}(q)$ is the FT of the renormalised chain
propagator. The bare parameters are expressed in terms of their renormalized
values by $ u_{0}=\mu^{\epsilon/2}S_{d}^{-1}Z_{u}u^{},$ $
t_{1}^{(0)}=\mu^{2}{Z_{\psi}^{-1}} \left[
  t_{2}^{}Z_{\psi,\phi^{2}}+t_{1}^{}Z_{\psi,\psi^{2}}\right]$ and $
t_{2}^{(0)}={\mu^{2}}{Z_{\phi}^{-1}} \left[
  t_{2}^{}Z_{\phi,\phi^{2}}+t_{1}^{}Z_{\phi,\psi^{2}}\right]$ where
$\mu^{-1}$ is an external length-scale and $S_{d}=\Omega_{d}/(2\pi)^{d}$ with
$\Omega_{d}$ the angular part of a $d$-dimensional integral.  We denote an $N$
point polymer $M$ point coulomb vertex function as $\Gamma^{(N,M)}$ while a
composite $N$ point polymer $M$ point coulomb vertex function with $L$,
$\phi^{2}$ and $K$, $\psi^{2}$ insertions is generally given by
$\Gamma^{(N,L;M,K)}$. The wave-function renormalization factors are calculated
by expressing the renormalized vertex functions in terms of the bare ones as
$\Gamma^{(N,M)}_{R}= Z_{{\phi}}^{N/2}Z_{{\psi}}^{M/2}\Gamma^{(N,M)}$.
The $Z$-factors are evaluated using the vertex functions which are calculated
using a dimensional regularisation scheme with minimal subtraction of
poles~\cite{ft}. We begin with the Coulomb vertex $\Gamma^{(0,2)}$ which due to
the $n \rightarrow 0$ limit gives the result $Z_{\psi}=1$ 
because all closed subdiagrams composed of these polymer lines give a factor
$n$.  From the polymer vertex $\Gamma^{(2,0)}$ we get $Z_{\phi}$.  For the
determination of $Z_{u}$ we need the leading singular $1/\epsilon$
contribution of $\Gamma^{(2,1)}$ from which follows the $\beta$-function the
zero of which yields the infra-red stable fixed point
$u*^{2}_{}=-{27\epsilon}/{4}$. An evaluation of $\Gamma^{{(0,0;2,1)}}$ yields
$Z_{\psi,\psi^{2}}=1$ and $\Gamma^{{(0,1;2,0)}}$ gives $Z_{\psi,\phi^{2}}=0$
using similar arguments as those for $Z_{\psi}$. The composite polymer vertex
function $\Gamma^{(2,1;0,0)}$ gives $Z_{\phi,\phi^{2}}$ and
$\Gamma^{(2,0;0,1)}$ to 1-loop yields $Z_{\phi,\psi^{2}}=Z_{\phi,\phi^{2}}-1$.

To see the scaling form of $\langle R^{2} \rangle$, we need to solve
the RG equation for ${\cal{G}}(q)\equiv{\cal T}_{{\cal L}}^{-1}[\tilde{G}_{R}^{(2,0)}(q)]$ which is given by \begin{equation}
\left( \mu \partial_{\mu} + \beta_{u}\partial_{u} - 2 t_{1}\partial_{t_{1}} -
  \vartheta_{2} {\cal L} \partial_{\cal L} + \gamma_{\phi}-\vartheta_{2} -
  \vartheta_{1}t_{1}{\cal L} \right) {\cal{G}}(q/\mu;u,{\cal
  L},t_{1})=0. \end{equation} where $\beta_{u}=(\partial u/ \partial \ln \mu)$,
 $\gamma_{\psi}=\beta_{u}\partial_{u} \ln Z_{\psi}$,
 $\gamma_{\phi}=\beta_{u}\partial_{u} \ln Z_{\phi}$,
 $\gamma_{\phi^{2}}=\beta_{u}\partial_{u} \ln Z_{\phi,\phi^{2}}$,
 $\vartheta_{1}= -Z_{\phi}/Z_{\phi,\phi^{2}}\beta_{u}\partial_{u}
 [Z_{\phi,\psi^{2}}/{Z_{\phi}}]$and $\vartheta_{2}= -2 - \gamma_{\phi^{2}} +
 \gamma_{\phi}$ and we have used the notation $\partial_{x}A =\partial
 A/\partial x$. The solution is readily obtained using the
method of characteristics~\cite{ft} giving 
\begin{eqnarray} F({\cal L},t_{1})&=&{\cal
    L}^{2\nu_{\phi}}F_{1}(1,t_{1}{\cal
    L}^{2\nu_{\phi}},u*)=1/t_{1}{F_{2}}({\cal
    L}t_{1}^{1/(2\nu_{\phi})},1,u*)\label{scalform} 
\end{eqnarray} showing non-trivial scaling
($\eta_{\phi}=\gamma_{\phi}(u*),\nu_{\phi}^{-1}-2=\gamma_{\phi^{2}}(u*)-\gamma_{\phi}(u*)$).
This means that the scaling function must be explicitly calculated to extract
information about the size of the chain.  Great care must be taken in the
interpretation of the inverse Laplace transform. As we are dealing with
critical fields, the masses $t_{1}$ and $t_{2}$ must go to zero. We get
non-trivial results only when {\it both} length-scales are large and of the
same order otherwise the larger length-scale dominates the physics and the
smaller becomes irrelevant.  This means that the ratio of the two
length-scales $c=\frac{\cal L}{1/\kappa^{2}} $ must be finite.  This gives the
non-trivial fixed-point $u* \neq 0$.  It is reassuring to note that the
explicit calculation, (\ref{rcool}), agrees {\it exactly} with the general
form of the scaling function in equation~(\ref{scalform}).
\begin{figure}
  \epsfxsize 8cm \rotatebox{-90}{\epsffile{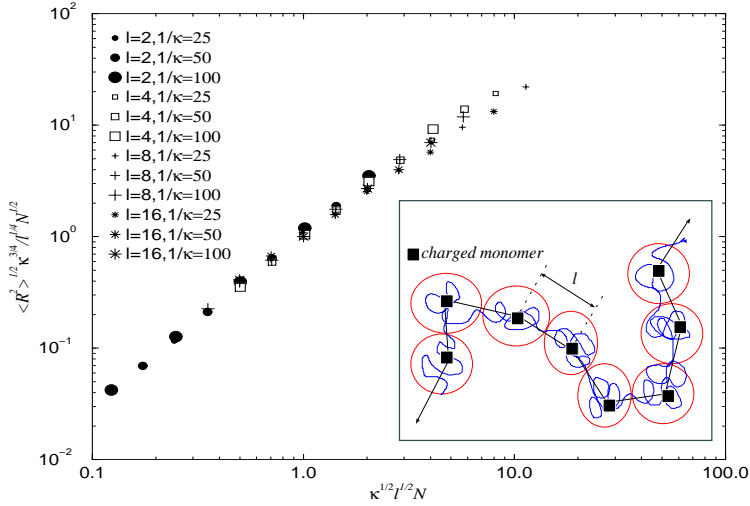}}
\caption{We plot Micka and Kremer's~\protect\cite{mickkrem} $3d$ simulation
  data of flexible DH chains for different values of $\ell$, $N$ and $\kappa$.
  $\sqrt{\langle R^{2}\rangle \bar{\kappa}^{2-1/\nu_{\phi}}/N}$ vs
  $\bar{\kappa}^{1/\nu_{\phi}}N$ with $\nu_{\phi}=2/(d-2)$ evaluated at $d=3$
  and we get {\it all} the points to collapse on one curve. Inset: weakly
  charged polyelectrolyte and Kuhn length $\ell$.}
\label{fig1}
\end{figure}

As is usual in RG analysis one must check for stability of the fixed point as
the quartic terms become relevant as we cross the dimension $d=4$. This says
something about the validity of the DH model (which assumes these operators
are irrelevant) in the physical dimension $d=3$. Any operators which make the
fixed-point unstable are {\bf corrections} to DH theory. The $\psi^{4}$ and
$\psi^{2}\phi^{2}$ and $(\phi^{2})^{2}$ are allowed by symmetry and become
relevant for $d<4$~\cite{Jug}. The excluded volume term $(\phi^{2})^{2}$ is
only relevant in the high salt regime. The quartic corrections mean that the
DH model is invalid below $d=4$ in the limit $\kappa \rightarrow 0$ because
the critical fluctuations of the Coulomb field for $d<4$ mean that it cannot
be described by a Gaussian model (no terms higher than quadratic in
$\psi$)~\cite{liu}.  Since we know that DH theory is in a sense a `mean-field'
approximation, we should expect it to be wrong in low dimensions.  Therefore
$\nu_{\phi}$ is valid for the DH model for $d<4$ but not for the {\it
  complete} theory (DH + corrections). The approximation that the fluctuations
in the density of the screening ions are small breaks down.  This does not
really have anything to do with $\nu_{\phi} > 1, d<4$ which is a {\bf result}
of a field theory of a DH chain.

Flexible polyelectrolytes have been simulated by
Stevens and Kremer~\cite{stevkrem} using explicit counter-ions and Micka and
Kremer~\cite{mickkrem} using a DH potential. Stevens and
Kremer~\cite{stevkrem} found exponents $\nu_{1} \simeq 0.94 <1$ in good
agreement with our picture. 
Micka and Kremer~\cite{mickkrem} observed  no unique
power law as $\kappa$ and $L$ were varied. In Fig.1 we plot their end-to-end
data using our explicit scaling form in equation(\ref{rcool}) and get very good
agreement. More detailed comparison of this and other quantities with their
numerical data will be presented elsewhere~\cite{solutions}.

In conclusion, we have systematically calculated the properties of a single
screened polyelectrolyte in the Debye-H\"uckel approximation (a well-defined
problem) and found non-trivial scaling behaviour in good agreement with
simulation.  The DH model is therefore controlled by {\it three} fixed points
governed by $c=\ell^{2}N\kappa^{2}$. They are $c\rightarrow\infty$ (SAW), $c
\simeq 1$ the new non-trivial behaviour which is the subject of this paper and
$c\rightarrow 0$ (rod). Our analysis confirms that the DH model breaks down if
there are large fluctuations in the coulomb field, as would be observed in
counter-ion condensation. We hope that this paper will stimulate the interest
in a 20 year old problem in the literature, our conclusion being that the
calculation of the exponents by Pfeuty et al and
others~\cite{Pfeuty,Jug,Mezard} was absolutely correct but their
interpretation of those exponents and hence of the physics was incomplete. We
believe that this paper therefore re-opens the area of the field theoretic RG
description of charged polymers.  In future publications we will deal with
charged manifolds, directed polyelectrolytes, solutions and
networks~\cite{solutions}.
%
%
\stars 

We acknowledge helpful discussions with M.E. Cates, S.F. Edwards, H-K.
Janssen, J-F. Joanny, M. Kardar, K. Kremer, A. Liu, U. Micka, K.
M\"uller-Nedebock, J. Rudnick, M. Schmidt, B. S\"oderberg, D. Thirumalai, T.
Vilgis and T. Witten. We thank L. Sch\"afer for pointing out
reference~\cite{Jug}. We particularly thank K. Kremer and T. Vilgis for a
critical reading of the manuscript.
%
%
%

\end{document}